\documentclass[preprint,aps,physrev,superscriptaddress,onecolumn,showkeys]{revtex4-2}

\usepackage{graphicx}
\usepackage{amsmath}
\usepackage{amssymb}
\usepackage{amsthm}

\usepackage[utf8]{inputenc}
\usepackage[T1]{fontenc}

\usepackage{notoccite}
\usepackage{esvect}

\usepackage{xcolor}
\usepackage[normalem]{ulem}	
	
\usepackage{float}

\usepackage{hyperref}
\hypersetup{
   bookmarksnumbered=true,
	colorlinks = true,    % farbige Links
	allcolors = blue,            % blaue Links fuer alles ausgenommen Papierversion
	breaklinks = true,   %link text darf ueber Zeilengrenzen gehen
	backref = true               % back springt zurueck zum Ursprung
}
\usepackage[all]{hypcap}

\let\titleoriginal\title           % save original \title macro
\renewcommand{\title}[1]{          % substitute for a new \title
    \titleoriginal{#1}%               % define the real title
    \newcommand{\thetitle}{#1}        % define \thetitle
}

\DeclareMathSymbol{\shortminus}{\mathbin}{AMSa}{"39} % Short minus symbol: $\shortminus$

\begin{document}

%\title{Challenges in Probing the $d$-Wave Spin Texture of Altermagnetic $\alpha$-MnTe by Spin-Resolved ARPES}
\title{Discerning ground state and photoemission-induced spin textures in altermagnetic $\alpha$-MnTe}
%---------------------------------------------------------------------------------------------------------

\author{D.A. Usanov*}
\affiliation{Institut de Physique, \'{E}cole Polytechnique F\'{e}d\'{e}rale de Lausanne, CH-1015 Lausanne, Switzerland}
\affiliation{Center for Photon Science, Paul Scherrer Institut, CH-5232 Villigen, Switzerland}

\author{S.W. D'Souza*}
\affiliation{University of West Bohemia, New Technologies Research Center, Plzen 30100, Czech Republic}

\author{A. Dal Din*}
\affiliation{School of Physics and Astronomy, University of Nottingham, Nottingham NG7 2RD, United Kingdom}

\author{J. Krempask\'y}
\affiliation{Center for Photon Science, Paul Scherrer Institut, CH-5232 Villigen, Switzerland}

\author{F. Guo}
%\affiliation{Center for Photon Science, Paul Scherrer Institut, CH-5232 Villigen, Switzerland}
\affiliation{Institut de Physique, \'{E}cole Polytechnique F\'{e}d\'{e}rale de Lausanne, CH-1015 Lausanne, Switzerland}

\author{O.J. Amin}
\affiliation{School of Physics and Astronomy, University of Nottingham, Nottingham NG7 2RD, United Kingdom}

\author{C. Polley}
\affiliation{MAX IV Laboratory, Lund University, 221 00 Lund, Sweden}

\author{M. Leandersson}
\affiliation{MAX IV Laboratory, Lund University, 221 00 Lund, Sweden}

\author{G. Carbone}
\affiliation{MAX IV Laboratory, Lund University, 221 00 Lund, Sweden}

\author{B. Thiagarajan}
\affiliation{MAX IV Laboratory, Lund University, 221 00 Lund, Sweden}

\author{T. Jungwirth}
\affiliation{Institute of Physics, Czech Academy of Sciences, Cukrovarnick a 10, 162 00 Praha 6 Czech Republic}
\affiliation{School of Physics and Astronomy, University of Nottingham, Nottingham NG7 2RD, United Kingdom}

\author{L. \v{S}mejkal}
\affiliation{Max Planck Institute for the Physics of Complex Systems, 01187 Dresden, Germany}
\affiliation{Max Planck Institute for Chemical Physics of Solids, 01187 Dresden, Germany}
\affiliation{Institute of Physics, Czech Academy of Sciences,
Cukrovarnick\'a 10, 162 00 Praha 6 Czech Republic}

\author{J. Min\'ar}
\affiliation{University of West Bohemia, New Technologies Research Center, Plzen 30100, Czech Republic}

\author{P. Wadley}
\affiliation{School of Physics and Astronomy, University of Nottingham, Nottingham NG7 2RD, United Kingdom}

\author{J.H. Dil}
\affiliation{Institut de Physique, \'{E}cole Polytechnique F\'{e}d\'{e}rale de Lausanne, CH-1015 Lausanne, Switzerland}
\affiliation{Center for Photon Science, Paul Scherrer Institut, CH-5232 Villigen, Switzerland}

\date{\today}% It is always \today, today,
             %  but any date may be explicitly specified

%Use showkeys class option if keyword
                              %display desired

%---------------------------------------------------------------------------------------------------------

\begin{abstract} 

The recently discovered class of altermagnets provide a physical realization of an unconventional compensated magnetic phase with a higher partial-wave type of ordering, reminiscent of unconventional superfluid phases. Their stability under normal conditions has sparked significant research interest, spanning fields from spintronics to topological and correlated quantum materials. Spin- and angle-resolved photoemission spectroscopy (SARPES) has great promise to resolve the momentum-dependent spin textures, which are intricately interweaved with the altermagnetic direct space spin order. Using the relativistic $d$-wave-like collinear spin polarization on one of the non-relativistic nodal surfaces of the altermagnetic band structure of $\alpha$-MnTe as an example, we here identify and resolve the challenges associated with (S)ARPES studies on altermagnets. We focus particularly on the role of photoemission-induced electron polarization and the coupling between light and the N\'eel vector of a magnetic domain. Our findings reveal an atypical behaviour of photoemission selection rules while using linearly-polarized light. Our methods allow to distinguish polarization of photoelectrons originating from the sample's ground state spin texture, on one hand, and from the photoemission process, on the other hand. Our experimental results are supported by a combination of ab initio band-structure and 1-step photoemission calculations.

%The results are supported and reproduced by 1-step photoemission calculations of a specific multidomain superposition, hence supporting the distorted $d$-wave out-of-plane spin texture of MnTe. 

\end{abstract}

%__________________________________________________________________________

\maketitle

%__________________________________________________________________________

\section{\label{sec:Introduction}Introduction}

Altermagnetism is a magnetic phase recently identified by a spin group theory and characterized by a compensated collinear order that breaks spin degeneracy in the electronic band structure. The unconventional nature of the altermagnetic ordering stems from the spontaneously broken spin-space and real-space rotation symmetries, while preserving a symmetry combining the spin-space and real-space rotation transformations \cite{vsmejkal2022beyond, vsmejkal2022emerging}. Unlike ferromagnetic ordering with nodeless $s$-wave-like spin polarization in the band structure generating net magnetization, and unlike  conventional antiferromagnetic ordering with spin-degenerate bands, the compensated altermagnetic ordering features a distinctive nodal $d$-wave ($g$ or $i$-wave) spin polarization of the band structure. \cite{vsmejkal2022beyond, vsmejkal2022emerging}. 

Since their theoretical conception, altermagnetic phases have been predicted in a wide variety of material classes, including insulators, semiconductors, semimetals, metals, and superconductors \cite{vsmejkal2022emerging, jungwirth2026symmetry, song2025altermagnets}. Among these candidates, manganese telluride (MnTe) stands out as a prototypical system that has attracted growing theoretical and experimental interest \cite{krempasky2024altermagnetic, osumi2024observation, lee2024broken, hajlaoui2024temperature, amin2024nanoscale, orlova2025magnetization, orlova2024crossover, belashchenko2025giant, lee2025dichotomous}. MnTe crystallizes in three main structural forms: zinc blende, wurtzite, and nickeline \cite{schlesinger1998mn, jain2024buffer}. The nickeline (NiAs-type) phase, known as $\alpha$-MnTe, is the focus of the present study. $\alpha$-MnTe is a semiconductor with an indirect band gap of approximately $E_g \approx 1.27$--$1.46$~eV \cite{bossini2020exchange}, and it belongs to the hexagonal space group $P6_3/mmc$. Its magnetic structure consists of two Mn sublattices, $A$ and $B$, aligned antiparallel with respect to each other with a N\'eel temperature of 310~K \cite{kriegner2016multiple}. These sublattices are connected by a non-symmorphic sixfold screw-axis along the $c$-axis, but are not related by simple translation or inversion \cite{krempasky2024altermagnetic, gonzalez2023spontaneous, vsmejkal2022beyond}. This symmetry-breaking arrangement gives rise to an electronic structure with non-relativistic $g$-wave spin splitting, characterized by three nodal planes $\Gamma$-K-A. The fourth nodal plane $\Gamma$-K-M (referred to as $k_z = 0$) arises from a mirror symmetry connecting the opposite-spin sublattices. This non-relativistic spin splitting has been previously referred as ‘strong’ altermagnetic lifted Kramer's spin degeneracy (LKSD) and discussed elsewhere \cite{krempasky2024altermagnetic}. 

\begin{figure}[h]
\centering
    \includegraphics[width=0.9\linewidth]{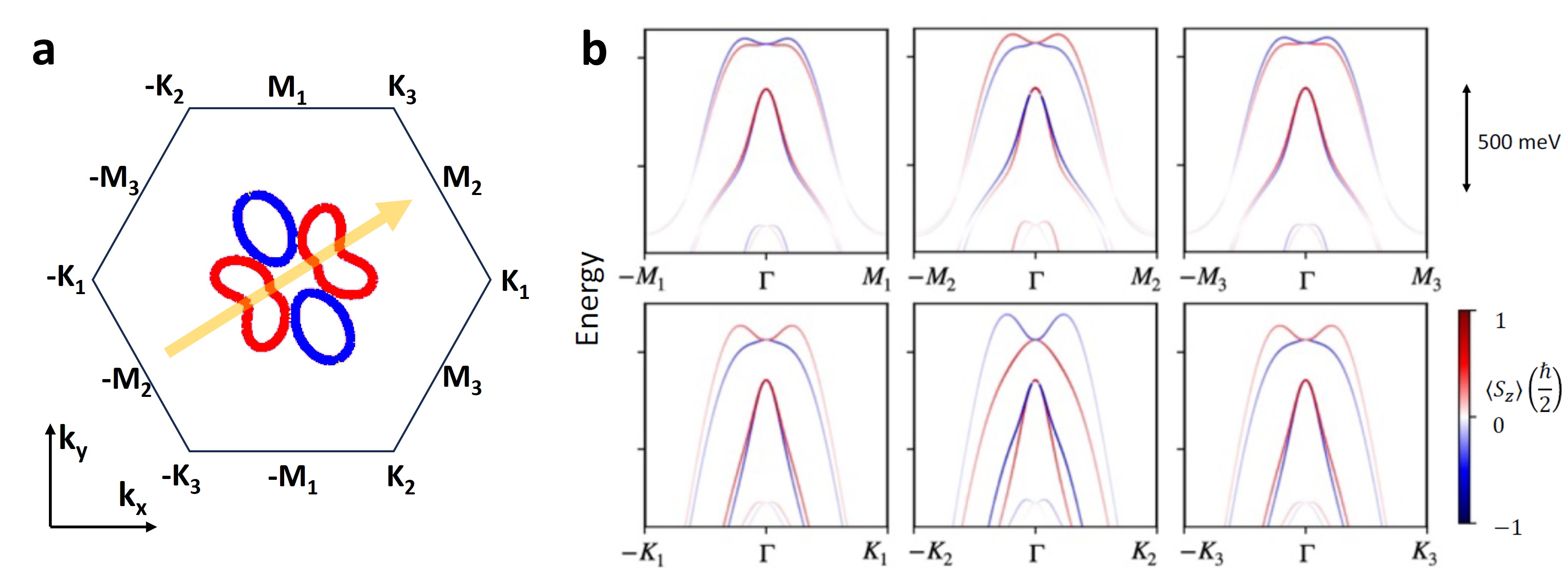}
    \caption{\label{figure: intro}(a) Schematic representation of 2D $d$-wave-like out-of-plane polarization symmetry of the reciprocal space. Notation of high-symmetry points of the Brillouin zone at $k_z = 0$ is shown. $d$-wave symmetry of $S_z$  is sketched on the Fermi surface cut at binding energy -0.1 eV. The N\'eel vector is indicated with a yellow arrow. (b) shows the spin-polarized band structure along the $\overline{\Gamma}-\overline{\text{K}}_i$, $\overline{\Gamma}-\overline{\text{M}}_i$ paths at $k_z = 0$ \cite{krempasky2024altermagnetic}.
    }
\end{figure}

The spin degeneracy within the nodal planes can be lifted by the spin-orbit coupling (SOC) in altermagnets without breaking the crystal inversion symmetry. In particular, within the $k_z=0$ nodal plane, the resulting spin polarization ($S_z$) is perpendicular to the axis of the in-plane manganese magnetic moments in the direct crystal space. This unconventional out-of-plane polarization component has a $d$-wave-like altermagnetic form (Fig. \ref{figure: intro}), resulting in different splitting magnitude and spin polarization along the high symmetry directions (Fig. \ref{figure: intro}b). The orientation of this $d$-wave-like spin splitting is dictated by the N\'eel vector, which has 6 possible orientations (below 250 K \cite{hicken2025anomalous}) along one of the easy axes (corresponding to the $\Gamma -M_i$ directions in the momentum space). This mechanism is called ‘weak’ altermagnetic LKSD \cite{krempasky2024altermagnetic, campos2026persistent} and is the main focus of this paper.

% therefore the splitting magnitude and polarization along high symmetry directions ($\Gamma$-K\textsubscript{i} and $\Gamma$-M\textsubscript{i}) depend on the N\'eel vector. There are two $\Gamma$-K\textsubscript{i}(-M\textsubscript{i}) paths showing the same spin splitting and one with a larger splitting and inverted polarization.
%Noteworthy, due to non-equivalence of $\Gamma$-K\textsubscript{i} and $\Gamma$-M\textsubscript{i} directions in a hexagonal crystal, this d-wave symmetry of $S_z$ is distorted. This mechanism is called ‘weak’ altermagnetic LKSD \cite{krempasky2024altermagnetic} and is the main focus of this paper.

Among experimental techniques used in studying altermagnetic materials (such as x-ray magnetic circular dichroism (XMCD) \cite{amin2024nanoscale, hariki2024x, takegami2025circular, kruger2025circular, lovesey2023templates, hariki2024determination} and anomalous Hall effect measurements \cite{vsmejkal2022anomalous, gonzalez2023spontaneous}), only spin- and angle-resolved photoemission (SARPES) \cite{ okuda2017recent, dil2019spin, schoenhense2022, din2025unconventional} is able to provide the most direct access to this unconventional $d$-wave-like out-of-plane spin polarization of the band structure. However, the angular distribution of photoelectrons is always affected by matrix element effects \cite{moser2017experimentalist}, which should be taken into account. Moreover, the complexity of ARPES data interpretation increases greatly when adding spin resolution, since electron polarization is also largely affected by the light polarization, experimental geometry and final states of photoemission. Matrix element effects are known to be able not only to alter polarization amplitude, but they may also completely change the observed spin texture with respect to the ground state \cite{heinzmann2012spin, meier2011interference, fanciulli2017spin, riley2014direct}. That is why it is essential to decouple signals arising from the ground state and from the photoemission process. 

This study aims to provide insights into challenges associated with (S)ARPES measurements of altermagnetic MnTe and data interpretation. We focus on the electronic bands close to the Fermi level and around the nodal plane $k_z = 0$ accessed from $(0001)$ surface of MnTe and consider only out-of-plane spin polarization $S_z$. Many of the considerations will also apply to other materials and spin directions, whereby the symmetry operations will be different and will depend on the expected spin quantization axis and experimental geometry.

%_____________________________________________
\section{\label{sec:Results and discussions}Results and discussions}

In order to easily visualize the spin textures, to controllably combine different domains, and to rapidly obtain a impression of the relevant parameters, we will start the discussion based on one-step photoemission calculations. These will subsequently be compared to corresponding ARPES results, with and without spin resolution. This avoids the need to perform time-demanding (S)ARPES experiments for a wide range of experimental parameters and allows us to focus on the relevant symmetry operations to discern photoemission induced spin signals for those related to the ground state. 

\subsection{\label{sec:1-step simulation of photoemission from MnTe}1-step simulation of photoemission from MnTe}

As discussed above, the band structure of a single magnetic domain in MnTe exhibits an out-of-plane spin texture of the $d$-wave-like symmetry around the $k_z = 0$ nodal plane. Both the sign and the orientation of this texture are determined by the in-plane N\'eel vector. However, in an (S)ARPES experiment, this ground-state spin texture is significantly influenced by matrix element effects of the photoemission process. These effects depend not only on the photon energy but also on the relative orientation of the sample, light polarization, and the general experimental geometry. In altermagnetic materials such as MnTe, the orientation of the N\'eel vector plays a particularly critical role. To illustrate this, let us consider one-step photoemission calculations at the $k_z = 0$ nodal plane using a photon energy of $h\nu \approx 80$~eV (see Fig.~\ref{figure: 1-step} and Fig.~\ref{figure: PZ MEE grid} in Appendix \ref{sec:Impact of matrix element effects on spin polarization of photoelectrons}). In these calculations, spin polarization maps of photoelectrons are computed for fixed experimental conditions, varying only the N\'eel vector orientation and/or the light polarization.

\begin{figure}[h]
\centering
    \includegraphics[width=0.95\linewidth]{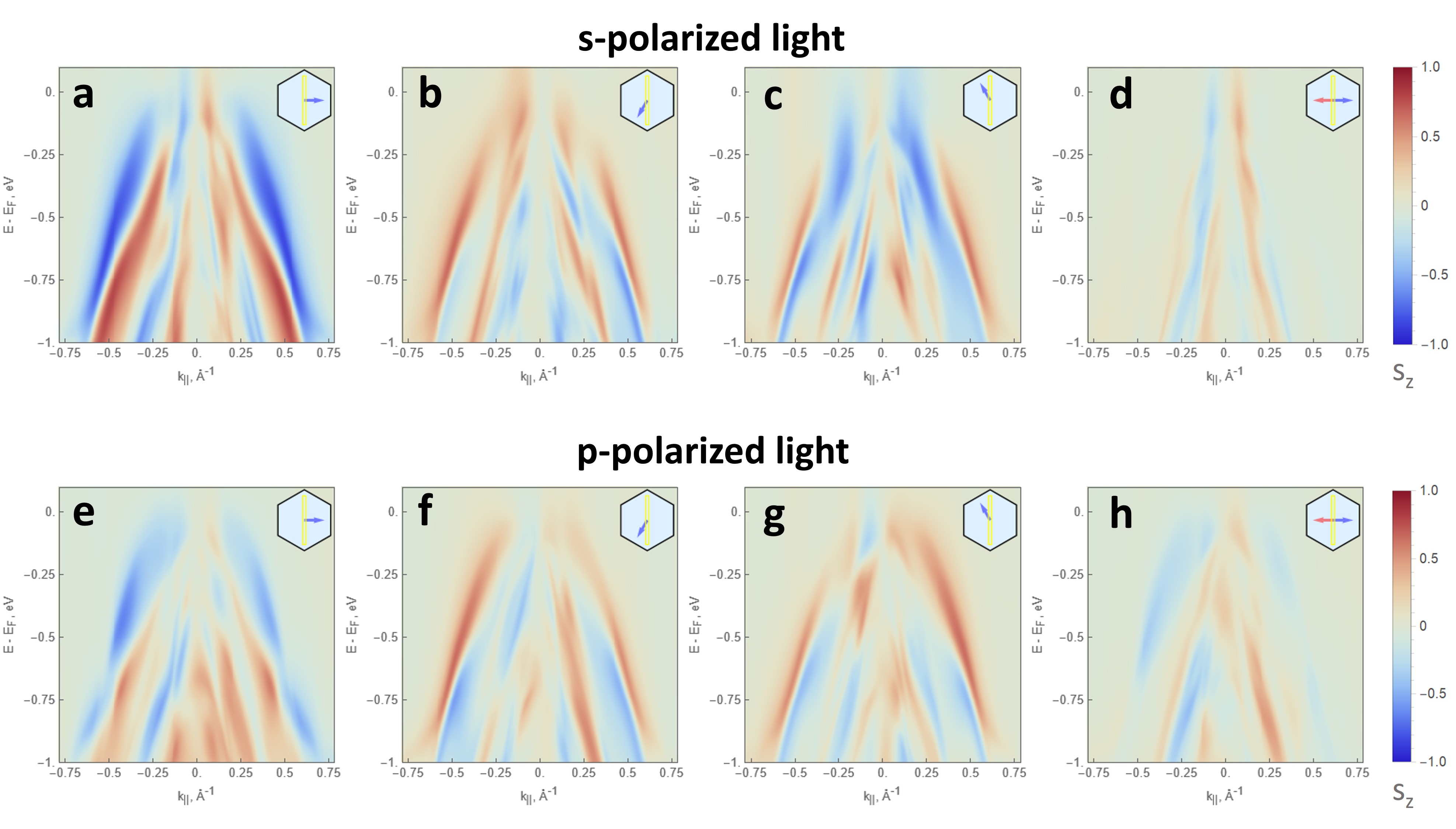}
    \caption{\label{figure: 1-step}Calculated SARPES maps of the out-of-plane spin polarization at the $k_z=0$ nodal plane at $h\nu=82$~eV along $\overline{\text{K}}-\overline{\Gamma}-\overline{\text{K}}$ direction. The domain orientation and experimental geometry in each plot are sketched in the inset. (a)-(d) correspond to the $s$-polarized light, (e)-(h) correspond to $p$-polarized light.}
\end{figure}

Across all configurations, the calculated spin polarization of the photocurrent reveals a structure more complex than that of the ground state. Specifically, and in line with the expected d-wave spin texture, the modulation of the out-of-plane spin component $S_z$ strongly depends on the orientation of the N\'eel vector relative to the ARPES measurement geometry. However, this modulation varies significantly when changing light polarization (Fig.~\ref{figure: 1-step}(e-h)). A similar dependence is observed in the spin-integrated photoemission intensity (see Fig. \ref{figure: 1-step SI} in Appendix \ref{sec:Comparison of the experimental and 1-step calculated data}), underscoring the critical role of matrix element effects in interpreting (S)ARPES data.

It is important to note that photoelectrons can posses non-zero spin polarization even if the ground state has effectively zero net polarization everywhere in the Brillouin zone. To illustrate this, we simulated photoemission from a sample consisting of two magnetic domains with opposite N\'eel vectors (Fig.~\ref{figure: 1-step}(d) and (h)). Since flipping of the N\'eel vector results in spin-reversal in the entire band structure, the summed ground-state spin polarization of such a sample becomes zero everywhere in the Brillouin zone. However, the resulting photocurrent polarization is clearly non-zero, with features that are symmetric and antisymmetric with respect to the $\Gamma$ point and with a strong dependence on the light polarization.

The influence of matrix element effects on spin polarization for all orientations of N\'eel domains is detailed in Appendix \ref{sec:Impact of matrix element effects on spin polarization of photoelectrons}. For the experimental geometry used in the presented (S)ARPES measurements, the use of $s$-polarized (LV) light adds a photoemission-induced spin signal, which is purely antisymmetric with respect to the $\Gamma$ point. In contrast, under $p$-polarized (LH) light, symmetric components emerge as well. Additionally, we identify a region in the Brillouin zone — defined by $k_{||} > 0.25$ and $E_B < 0.5$~eV — where the impact of matrix element effects is minimal. This region-of-interest (ROI) is therefore well suited for probing the intrinsic spin texture of the ground state and has been predominantly used in our SARPES measurements of patterned thin films \cite{din2025unconventional}.

Based on the one-step calculations, we propose a simple rule for interpreting out-of-plane spin polarization signals around the $k_z = 0$ nodal plane at $h\nu \approx 80$ eV in SARPES experiments of $(0001)$ MnTe surface. Any $S_z$ spin polarization component that is antisymmetric with respect to the $\Gamma$ point originates exclusively from matrix element effects. In contrast, symmetric components may reflect the symmetry of both the ground-state spin texture and matrix element contributions. However, in the special case of $s$-polarized light, the symmetric part of the spin signal reflects the ground-state properties.

Another important aspect is related to the specific symmetry of the MnTe crystal structure. Non-symmorphic space groups, characterized by the existence of non-primitive symmetry operations such as screw axes and glide planes, significantly complicate the interpretation of ARPES data. Unlike symmorphic crystals, where the symmetry of the observable final state is uniquely defined by the direction of its wave vector, non-symmorphicity renders this final state symmetry non-unique, as it becomes a function of both the direction and the length (magnitude) of the wave vector \cite{pescia1985determination}. This leads to symmetry switching at Brillouin zone boundaries, where the irreducible representation of the final state plane wave alternates between consequent zones. A primary experimental consequence of this switching is an apparent doubling of the Brillouin zone dimension along specific symmetry lines, as transitions from certain initial bands may be dipole-forbidden in an even zone but allowed in an odd one \cite{prince1986symmetry, arpiainen2006effect, matsui20184}. The associated photocurrent modulation has been observed experimentally with spin-integrated ARPES in the soft X-ray  \cite{krempasky2024altermagnetic} and and VUV \cite{din2025unconventional} ranges. Furthermore, non-symmorphisity imposes momentum-dependent spin selection rules, causing the spin polarization of photoelectrons from a single initial state to reverse, when their momentum moves into adjacent Brillouin zones \cite{ryoo2018momentum, alexandradinata2020glide} along directions associated with non-symmorphic operations.

Regarding MnTe, we hypothesize a similar effect as reported for non-symmorphic space groups with glide planes \cite{ryoo2018momentum, alexandradinata2020glide}. In particular, the presence of the 6-fold screw axis along the $[0001]$ direction may lead to an asynchronous intensity modulation of both $S_z$ spin channels when changing the photon energy. We therefore emphasize the importance of symmetry arguments, and not necessarily the magnitude nor absolute sign of photoelectron spin polarization when interpreting SARPES data.

%______________________________________________________________
\subsection{\label{sec:Influence of multiple domains}Influence of multiple domains}

(S)ARPES measurements of real $\alpha$-MnTe samples are, in fact, even more complex than the 1-step predictions described above. This is associated with the fact that the characteristic size of a magnetic domain in MnTe thin films is of the order of 0.1 $\mu m$ in the absence of patterning \cite{amin2024nanoscale}. Therefore, any conventional synchrotron-based ARPES experiment on unpatterned films will most likely probe simultaneously several domains with different orientation. 

In $\alpha$-MnTe with the surface along the $[0001]$-direction, there are 6 possible N\'eel vector orientations, given the three equivalent $\Gamma$-M\textsubscript{i} easy-axes. For a given easy axis, the band structure of domains with the opposite N\'eel vector are spin-flipped with respect to each other. In general, we can define two basic scenarios of possible domain configurations under the photon beam: either the overall domain composition is balanced or there is a preponderance of a certain easy axis and sign of the N\'eel vector. The first case is logically assumed for a non-field-cooled unpatterned film, where the total net spin polarization in the ground state averages out to zero everywhere in the Brillouin zone. We will discuss the balanced case first, and will focus on imbalanced domain configurations afterwards.

%____________________________________________
\subsubsection{\label{sec:Balanced domain configuration}Balanced domain configuration}

\begin{figure}[h]
\centering
    \includegraphics[width=0.95\linewidth]{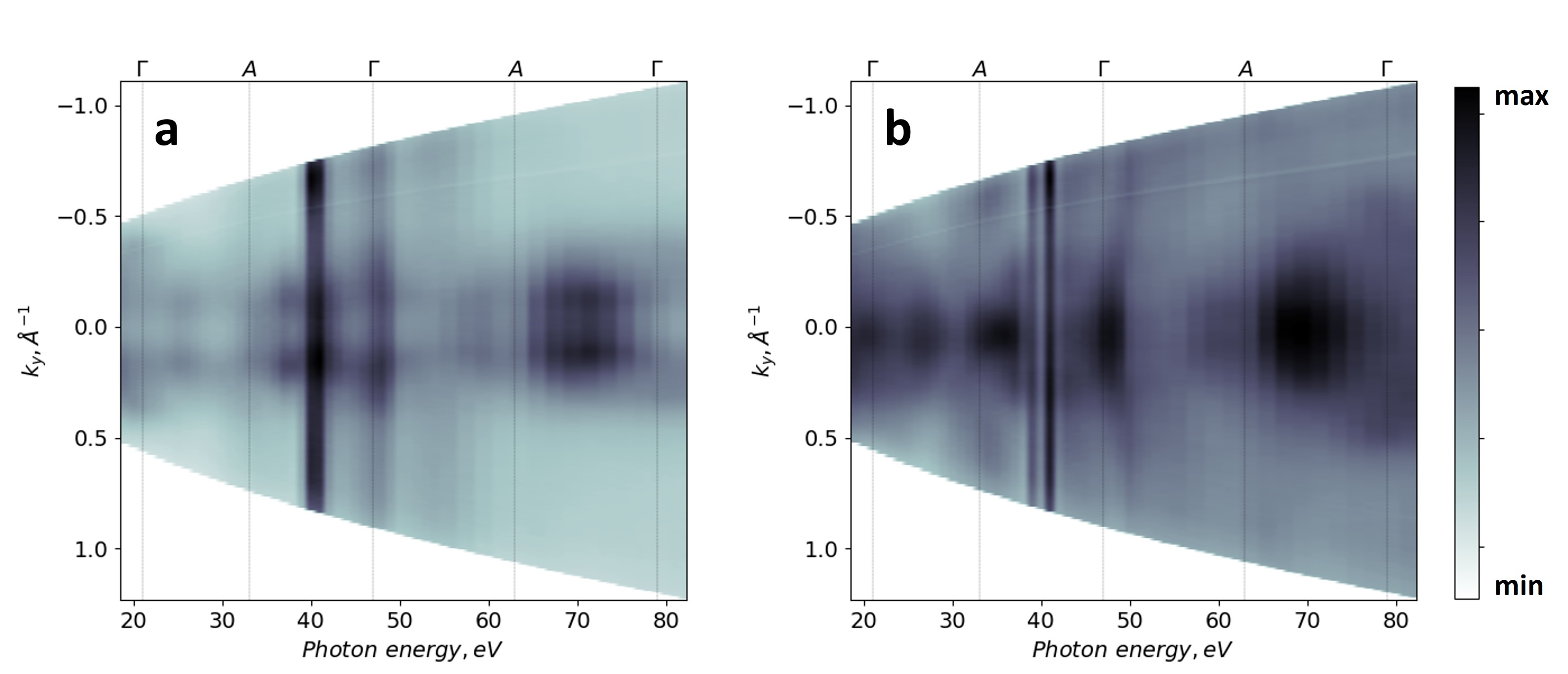}
    \caption{\label{figure: hv scan}Constant energy slices (100 meV integration) from the photon energy scan at (a) $E_B = 150~meV$ and (b) $E_B = 500~meV$ measured with $p$-polarized light along the $\overline{\text{K}}-\overline{\Gamma}-\overline{\text{K}}$ direction. Positions of high symmetry points for the normal emission are indicated. Constant-angle intense lines at $h\nu \approx 40$ and $48$ eV are core levels excited due to higher order harmonics coming from the undulator.}
\end{figure}

We have preformed ARPES measurements on non-field-cooled unpatterned films in the VUV range with and without spin resolution. First, photon energy, and thus $k_z$ scans in the range between $h\nu=19$ to $82$~eV were measured in order to locate the $k_z=0$ nodal planes. In Fig. \ref{figure: hv scan} constant energy slices through the specified photon energy range are displayed at binding energies of 150 and 500 meV. From the observed periodicity, and comparison to $ab-initio$ calculations and already published data \cite{hajlaoui2024temperature, lee2025dichotomous, osumi2024observation, lee2024investigations, martuza2025itinerant}, several $k_z=0$ planes can be located. For our further investigation we choose the one around $h\nu=78$~eV because of the clean spectral weight and the large signal to noise ratio over a large binding energy range. Importantly, in this photon energy range we mainly observe bands dispersing in $k_z$, thus indicating their bulk origin. At the same time, we do not exclude a presence of a surface resonance close to the zone center at around $k_{||} < \pm 0.2 $\AA\textsuperscript{-1} visible in Fig. \ref{figure: hv scan}a. 

\begin{figure}[h]
\centering
    \includegraphics[width=0.95\linewidth]{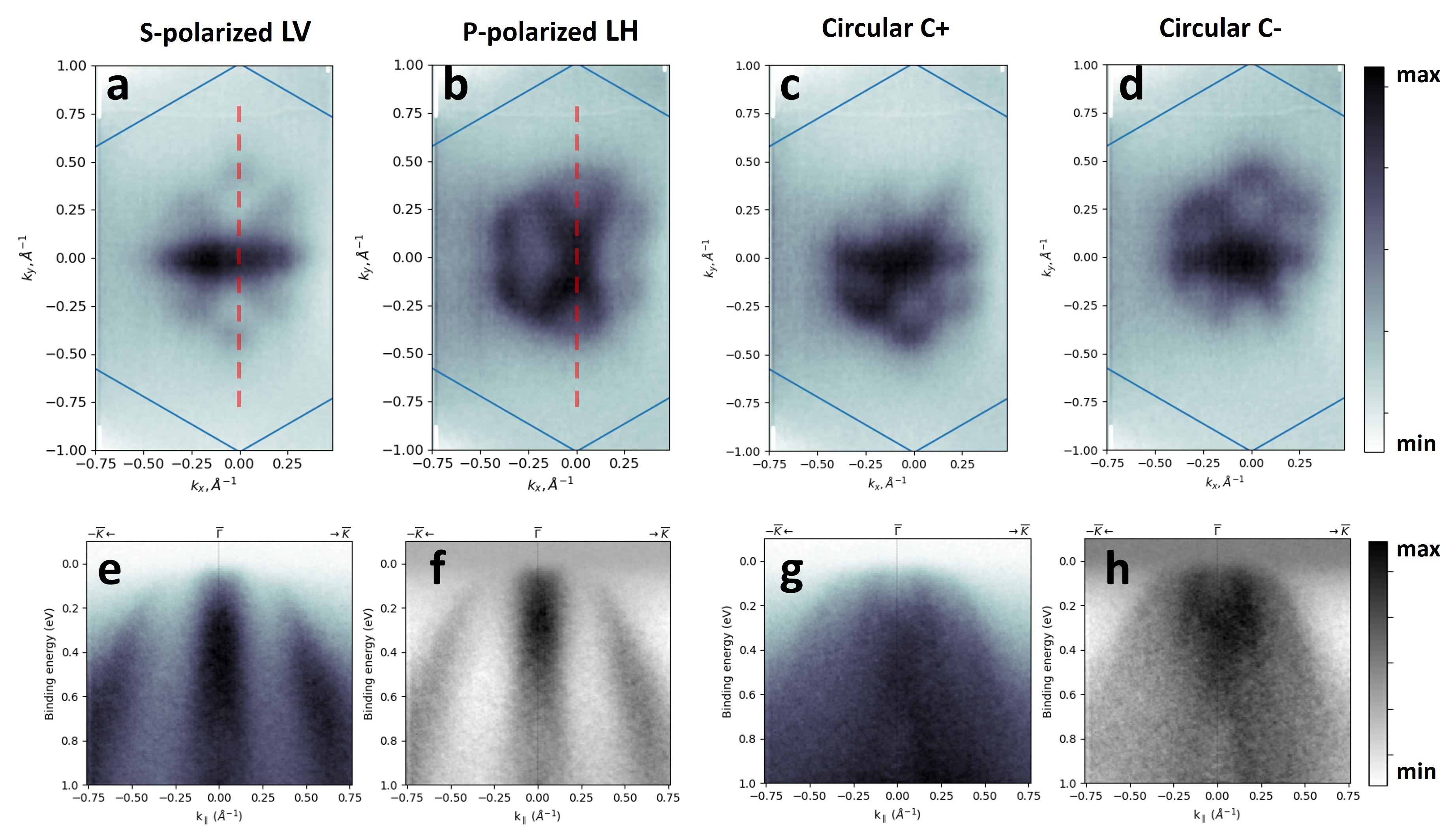}
    \caption{\label{figure: ARPES}(a)-(d) Constant energy maps 100 meV below $E_F$ for 4 different light polarizations, taken with $h\nu=78$ eV, the borders of the surface Brillouin zone are indicated; (e)-(h) corresponding bandmap cuts along the $\overline{\text{K}}-\overline{\Gamma}-\overline{\text{K}}$ directions, indicated with dashed red lines in (a) and (b) obtained with s-polarized (e,f) and p-polarized light (g,h); (e) and (g) show the raw data, (f) and (h) display the data with angle-integrated background subtracted for better visibility.}
\end{figure}

In Figure \ref{figure: ARPES} we plot constant energy surfaces 100~meV below the valence band maximum and the band dispersion along the $\overline{\text{K}}-\overline{\Gamma}-\overline{\text{K}}$ direction for different light polarizations near the $k_z=0$ nodal plane at $h\nu=78$.% and $h\nu=21$ eV. 
Similar to previous works \cite{krempasky2024altermagnetic, lee2024broken, osumi2024observation, hajlaoui2024temperature}, a clear band splitting on the order of 100~meV can be observed along the $\overline{\Gamma}-\overline{\text{K}}$ direction. Along $\overline{\Gamma}-\overline{\text{M}}$ the splitting is smaller and hence more difficult to resolve (Fig. \ref{figure: intro}). 

Both linear and circular dichroism are clearly visible in the constant energy surfaces. Because the latter is antisymmetric with regard to the scattering plane (which coincides with a mirror plane), we ascribe it to an effect of the experimental geometry \cite{schonhense1990circular} (Fig.~\ref{figure: CDAD}(b) in Appendix~\ref{sec:Circular dichroism in ARPES data}). In turn, the iso-energy cuts obtained by summing both circular polarizations (see Fig.~\ref{figure: CDAD}(a) in Appendix~\ref{sec:Circular dichroism in ARPES data}) effectively suppress the spectral weight modulation due to selection rules. Because of the finite energy, $k_z$ and angular resolution of the ARPES experiment compared to the difference in spin splitting between nonequivalent high-symmetry directions, the observed shape of these iso-surfaces is consistently 6-fold symmetric. Therefore, it suggests that spin-integrated ARPES from the $[0001]$ surface of MnTe is much less sensitive to the domain orientation than spin-resolved ARPES.

%Because we do not observe 2-fold signatures of a single domain, it does not seem possible to deduct the exact domain configuration.

At the same time, the linear dichroism and the corresponding spectral weight distribution (Fig. \ref{figure: ARPES}(a-b)) can not be explained solely by standard dipole selection rules \cite{moser2017experimentalist}. The angular distribution of photoelectrons observed with $p$-polarized light does not contradict the dipole selection rules of photoemission from $p_z$ orbitals (since these Te bands near $E_F$ have the dominant $p$-orbital component along the c-axis \cite{krempasky2024altermagnetic}). However, $s$-polarized light should result in an intensity minimum within the scattering plane (for all $p$ and $d$ orbital components), which is exactly the opposite with respect to our experimental observation. Apart from a violation of dipole selection rules and the plane-wave final state picture of photoemission, we hypothesize a selective photoexcitation of different N\'eel domains by linearly-polarized light. Spectral weight is enhanced within the scattering plane, when $\vv{E}$ is perpendicular to it, emphasizing mostly domains with the N\'eel vector perpendicular to $\vv{E}$. As will be explained below, this selectivity also affects the observed spin polarization of photoelectrons.

In Figure \ref{figure: SARPES}(a) we present the calculated spin signal along $\overline{\text{K}}-\overline{\Gamma}-\overline{\text{K}}$ for the configuration with six equally contributing domains and p-polarized light. Contrary to what would be expected from the ground state spin texture, a clear antisymmetric spin signal is present close to the zone center. As discussed above, this spin polarization originates purely from the selection rules of the photoemission process and is also visible close to the Fermi level in the experimental SARPES data obtained for small in-plane momenta ($k_{||}\approx 0.2~\text{\AA}^{-1}$) in the lower panel of Figure \ref{figure: SARPES}(a). This antisymmetric spin texture for a balanced domain structure when using p-polarised light, explains the Rashba-like spin signal obtained for MnTe single crystals under these conditions \cite{zeng2025non}. For larger momenta, the spin signal is expected to be practically zero in our case of a balanced domain configuration and $h\nu\approx 80$, which is related to the spin-compensated polarization of the ground state. 

%_____________________________________________________
\subsubsection{\label{sec:Imbalanced domain configuration}Imbalanced domain configuration}

In order to obtain access to ground-state-related spin polarization, an imbalance in the domain population should be created. In our case, we employed field cooling - cooling down the sample from a paramagnetic state in presence of static out-of-plane magnetic field \cite{amin2024nanoscale}. In this section we will present the results obtained from a field-cooled thin film of MnTe. 

\begin{figure}[h]
\centering
    \includegraphics[width=0.9\linewidth]{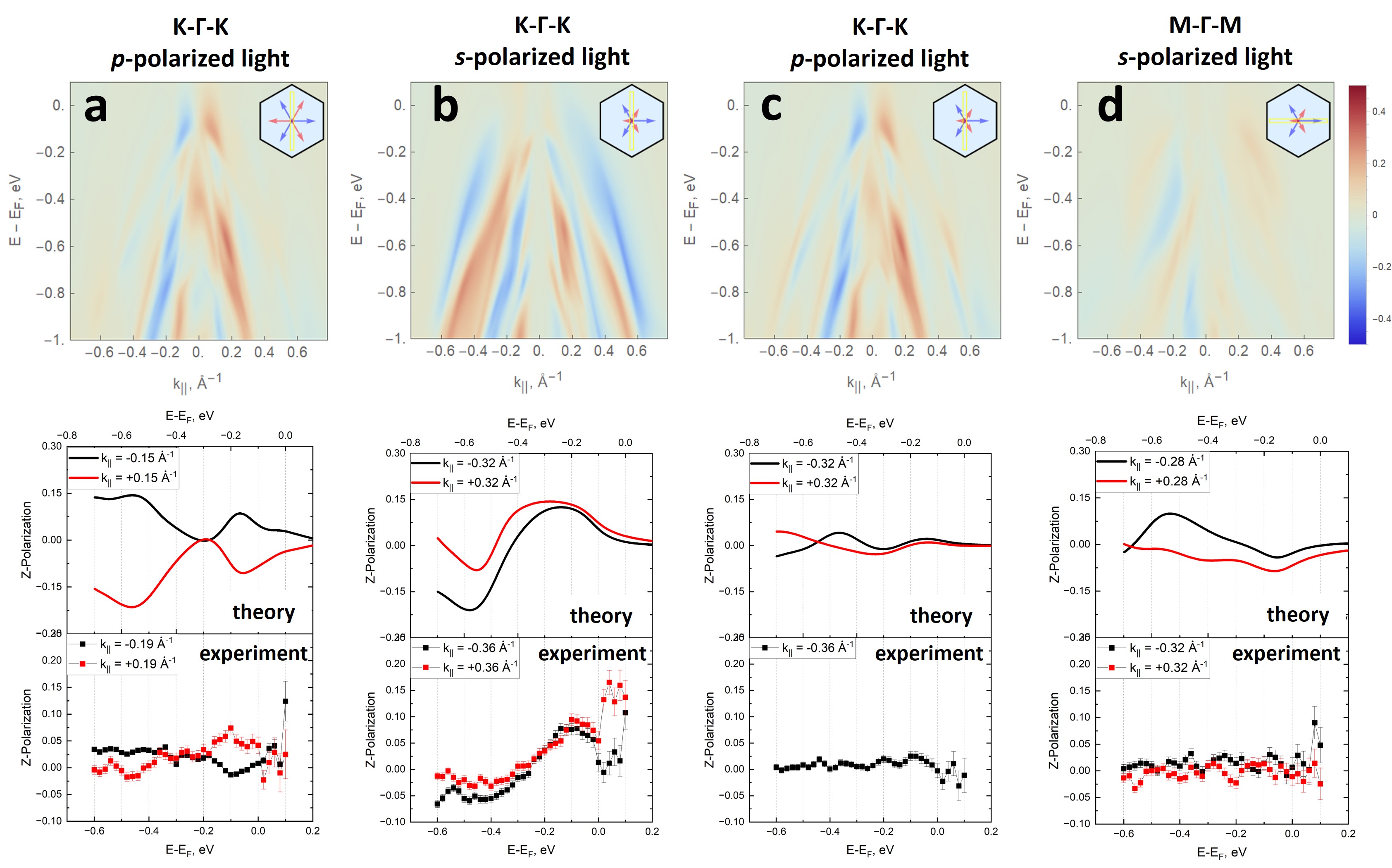}
    \caption{\label{figure: SARPES}SARPES 1-step calculations and measurements for the balanced (a) and imbalanced (b)-(d) domain configurations. Top panels show summed calculated band maps, middle panels EDCs extracted from these maps, and lower panels corresponding experimental data. The measured data were taken at $h\nu=78$ eV and the calculated ones at $h\nu=82$ eV. The experimental geometry and domain combinations are shown in the insets.}
\end{figure}

In an experiment on thin films, which have not undergone any micropatterning, we do not expect to get a single domain after the field cooling. Instead, we expect to induce an imbalance only between domains with opposite sign of the N\'eel vector \cite{amin2024nanoscale}. In Figure \ref{figure: SARPES}(b-d) such an imbalance is simulated for a domain configuration sketched in the inset and has the following composition: Ground state 1 (blue arrows) = (24; 19; 19)\%, Ground state 2 (red arrows) = (10; 14; 14)\%, corresponding to an average domain imbalance (effective "magnetization") of 0.28. Besides the antisymmetric spin signal close to $\Gamma$, now a clear symmetric spin texture, originating from the ground state, is expected at higher momenta. This is exactly what has been observed experimentally in a SARPES measurement of the field-cooled sample using s-polarized light. Spin-resolved energy-distribution curves (EDC) at $k_{||}\approx 0.36~\text{\AA}^{-1}$ (lowest panel) show a clear symmetric spin signal, resembling the EDCs extracted from the calculations (middle panel). Spin-resolved EDCs at higher momenta possess the same symmetry (see Fig.~\ref{figure: SARPES SI} in Appendix~\ref{sec:Supplementary SARPES data}). Since the impact of the matrix element effect has been predicted to be minimal in this region of the Brillouin zone (Fig.~\ref{figure: PZ MEE grid} in Appendix~\ref{sec:Impact of matrix element effects on spin polarization of photoelectrons}), the observed polarization signal reflects mostly the ground-state spin texture. Therefore, these data support the concept of the SOC-induced lifted Kramers spin degeneracy in altermagnetic MnTe. 

Based on the 1-step photoemission calculations, we summarize the effect of domain orientation on the measured spin polarization of photoelectrons excited by linearly-polarized light (a more detailed description is available in \cite{usanov2025}). Let us first consider the $\overline{\text{K}}-\overline{\Gamma}-\overline{\text{K}}$ direction perpendicular to the light scattering plane and utilizing $s$-polarized light (Fig. \ref{figure: SARPES}(b)). Our analysis of the 1-step photoemission calculations shows that mainly the domains with the N\'eel vector parallel to the scattering plane define the polarization texture in the ROI; all other domains provide smaller photoelectron current. Consequently, the polarization signal in this region of the Brillouin zone mainly reflects an imbalance between two domains of opposite N\'eel vectors perpendicular to $\vv{E}$ (Fig. \ref{figure: 1-step}(d)). Spin-resolved EDCs are predicted to be symmetric with respect to the $\Gamma$ point with a small impact of antisymmetric features. This was indeed observed in the experiment (Fig. \ref{figure: SARPES}(b) and Fig.~\ref{figure: SARPES SI} in Appendix~\ref{sec:Supplementary SARPES data}). Moreover, even for a perfect balance of magnetic domains with their N\'eel vector oriented 120$^o$ with respect to each over, the symmetric character of $S_z$ in ROI remains. 

A completely different picture is formed by utilizing $p$-polarized light with the same experimental geometry and domain configuration (Fig. \ref{figure: SARPES}(c)). Features related to the ground state have almost completely disappeared. What remains is mostly anti-symmetric (with respect to the zone center) and has a predominant photoemission origin. Specifically for the ROI defined above, the polarization signal is negligibly small and hence less feasible to detect in a real SARPES experiment. This was confirmed by measuring a spin-resolved EDC on one side at $k_{||} = -0.36$ \AA\textsuperscript{-1}, where polarization does not exceed a few \% and is significantly smaller as what was found for s-polarized light in panel (b).

Regarding the $\overline{\text{M}}-\overline{\Gamma}-\overline{\text{M}}$ direction, which lies within the scattering plane, and photoemission using $s$-polarized light (Fig. \ref{figure: SARPES}(d)), the domain superposition does not generally lead to a very distinguishable polarization signal, in contrast to $\overline{\text{K}}-\overline{\Gamma}-\overline{\text{K}}$. Even though both symmetric and anti-symmetric spin signals are expected in the ROI, they lack a general trend (in contrast to $\overline{\text{K}}-\overline{\Gamma}-\overline{\text{K}}$) and their amplitude is comparable within the detection limit. This model qualitatively reproduces the measured polarization, where no clear spin signal has been detected.

%_______________________
\section{\label{sec:Conclusions}Conclusions}

In this work, we presented a combined numerical and experimental (S)ARPES investigation of the SOC- and altermagnetism-induced spin texture in $\alpha$-MnTe, addressing both the experimental challenges of spin-resolved photoemission measurements and the complexities inherent in interpreting the resulting data.

(S)ARPES measurements have been performed on the $(0001)$ surface of MnTe thin films around the $k_z = 0$ nodal plane at $h\nu=78$~eV. Linear dichroism effects, that go beyond simple dipole selection rules, suggest a selective excitation of magnetic domains, depending on the orientation of the N\'eel vector with respect to both the light polarization and the scattering geometry. This coupling between the light polarization and the N\'eel vector was further corroborated by SARPES measurements on a field-cooled sample: a clear spin polarization signal was observed with $s$-polarized light, whereas $p$-polarized light resulted in an effectively unpolarized signal. Supported by 1-step photoemission calculations, the ground-state-related polarization signal has been decoupled from the polarization induced by the photoemission process. In particular, under specific experimental conditions, spin polarization, which is symmetric with respect to the $\Gamma$ point, is attributed to the ground-state spin texture, whereas antisymmetric polarization originates from the photoemission process. These findings are consistent with a qualitative model involving a superposition of multiple domains, each exhibiting a $d$-wave-like spin polarization.

Our results highlight the intrinsic complexity of SARPES investigations in MnTe and in altermagnetic materials in general. First, the band structure with lifted Kramers degeneracy - including both occupied and unoccupied states — introduces significant photon-energy-dependent final-state effects. Second, the interaction of spin-polarized bands with polarized light causes additional modulations of photoelectron current and spin polarization. Third, these effects are further influenced by crystal orientation, the direction of the N\'eel vector, and the experimental geometry. These considerations have to be taken into account when aiming to assess properties of the initial state spin texture. In combination with symmetry arguments, it enhances the feasibility to support the $d$-wave-like concept in MnTe experimentally \cite{din2025unconventional}. The final degree of complexity arises from the presence of multiple magnetic domains within the ARPES probing area. All of these factors must be carefully considered to reliably interpret spin-resolved photoemission data in altermagnetic systems to avoid interpreting photoemission induced spin polarization as representing the ground state. As a general rule of thumb, symmetry arguments based, for example, on N\'eel vector reversal or rotation, are more reliable in their interpretation as those based on changing photon energy or polarization, or on absolute sign or direction of the measured spin polarization. Lastly, our results show the importance of numerical experiments for the interpretation of complex data from altermagnetic systems.

%___________________________________________________________
\section{\label{sec:Methods}Methods}

The $\alpha$-MnTe thin films were synthesized at the University of Nottingham, with a thickness of 30 nm, grown at 700 K by molecular beam epitaxy (MBE) on InP(111)A substrates with c-axis aligned with the  surface normal, analogous to \cite{amin2024nanoscale}. The samples were transferred in a custom-build vacuum suitcase under UHV conditions ($p\approx 8\times10^{-11}$\ mbar) to the end-station without breaking the vacuum. Field cooling has been performed using permanent magnets (0.4 T field) with the magnetic field aligned with the $(0001)$ orientation of the films. The samples were first heated to 350 K (above the transition temperature of around 310 K), then cooled down to a temperature of 200 K using liquid nitrogen, before transfer and cool down to 80 K. P-type doping of the films has been identified \cite{amin2024nanoscale}.

The electronic structure calculations shown in Fig. \ref{figure: intro} were performed for bulk MnTe (space group $P6_3/mmc$, No. 194). The lattice parameters were taken from X-ray diffraction measurements \cite{kriegner2017magnetic}. Electronic structure calculations were carried out using the Vienna Ab initio Simulation Package (VASP) \cite{kresse1996efficient}, employing the Perdew–Burke–Ernzerhof (PBE) functional \cite{perdew1996generalized} with spin–orbit coupling enabled and electron correlations treated using the Dudarev-type spherically symmetric Hubbard correction as in Ref. \cite{gonzalez2023spontaneous}. An 8 × 8 × 5 k-point mesh and an energy cutoff of 520 eV were used.

The spin and angle-resolved photoemission (SARPES) calculations are preformed by employing one-step model of photoemission, originally proposed by Pendry and co-workers \cite{pendry1976theory, hopkinson1980calculation, pendry1974low}, as implemented in the fully relativistic SPR-KKR program package \cite{ebert2012munich, Ebert_2011}. Within the framework of multiple scattering theory, photocurrent from the Mn-terminated surface of hexagonal MnTe(0001) has been calculated in a direct way by determining the initial and final states for a semi-infinite atomic half-space using the low-energy electron diffraction (LEED) method \cite{gray2011probing}. The surface potential is described through a spin-dependent Rundgren–Malmstr\"om barrier \cite{malmstrom1980program}. The On-site Coulomb interaction parameters for Mn 3d states ($U=4.80eV$, $J=0.80eV$) within LSDA+U were chosen. The final states are calculated as the best available single-particle approach, “time-reversed spin-polarized LEED state”, and initial states are represented by the retarded one-electron Green function in a semi-infinite half-space. The photoemission calculations are performed by taking into consideration the actual measurement ARPES geometry to correctly account for the wave-vector, spin- and energy-dependent transition matrix elements \cite{pendry1974low, braun1996theory, braun2018correlation} along with all multiple scattering effects in the initial and final states, the effect of the photon momentum vector, and the escape depth of the photoelectrons via an imaginary part in the potential function.

In order to simulate the photoemission current from a particular domain configuration, signal from each domain orientation and for a selected light polarization in the common emission plane are added incoherently. Spin-resolved energy distribution curves in Fig. \ref{figure: SARPES} were extracted from the 1-step calculations taking into account finite energy and $k_{||}$ resolution of the experiment.

Spin-integrated and spin-resolved measurements on MnTe films have been performed at the B-branch of the Bloch beamline \cite{polley2024bloch} in the range of $h\nu$ = 19-82 eV at 80 K. The beam spot size of the sample is expected to be around 12.5 ~$\mu$m x 15.5 $\mu$m. For the presented spin-resolved data we used an energy resolution of 57 meV and angular integration of 1.5$^{\circ}$ along the analyzer slit and 0.5$^{\circ}$ perpendicular to it, corresponding to approx. 0.1 \AA\textsuperscript{-1} and 0.04 \AA\textsuperscript{-1} at $E_{kin} = 73$~eV respectively. A VLEED (Very Low Energy Electron Diffraction) spin detector (Ferrum\texttrademark~from Focus GmbH) was employed as electron polarimeter and is based on the spin-dependent exchange scattering of low energy electrons from a magnetized target (oxidized 10nm-thick Fe film on a W(001) crystal). 

In all measurements and calculations, the c-axis of the MnTe crystal structure was perpendicular to the sample surface, exposing either a manganese or tellurium layer. All spin-resolved data presented in this paper correspond to the $S_z$ polarization component aligned with the c-axis of the MnTe crystal. The c-axis was aligned with the optical axes of the hemispherical analyzer. The  $\overline{\text{K}}-\overline{\Gamma}-\overline{\text{K}}$ direction was aligned with the HSA entrance slit.

\section{\label{sec:Acknowledgments}Acknowledgments}

D.A.U. acknowledges support from the Swiss National Science Foundation through the Spark grant CRSK-2 228962.  F.G. and J.H.D. acknowledge support from the Swiss National Science Foundation (SNSF) Project No. 200021-200362. S.W.D and J.M acknowledge support from the QM4ST project funded by Programme Johannes Amos Commenius, call Excellent Research (Project No. CZ.02.01.01/00/22\_008/0004572). LS acknowledges support from the ERC Starting Grant No. 101165122, the ERC Advanced Grant no. 101095925. O.J.A. acknowledges support from the Leverhulme Trust Grant ECF-2023-755. P.W. acknowledges support from the Royal Society through a University Research Fellowship. A.D.D acknowledges support from the EPSRC grant EP/V031201/1. T.J acknowledges support from ERC Advanced Grant 101095925. We thank MAX IV Laboratory for time on Beamline BLOCH under proposal 20231587. The authors greatly appreciate theoretical discussions with Eugene Krasovskii and Christophe Berthod.  

\section*{Data Availability Statement}

The data that support the findings of this study are available from the corresponding author upon reasonable request.

\appendix
\section{\label{sec:Comparison of the experimental and 1-step calculated data}Comparison of the experimental and 1-step calculated data}

\begin{figure}[h]
\centering
    \includegraphics[width=0.95\linewidth]{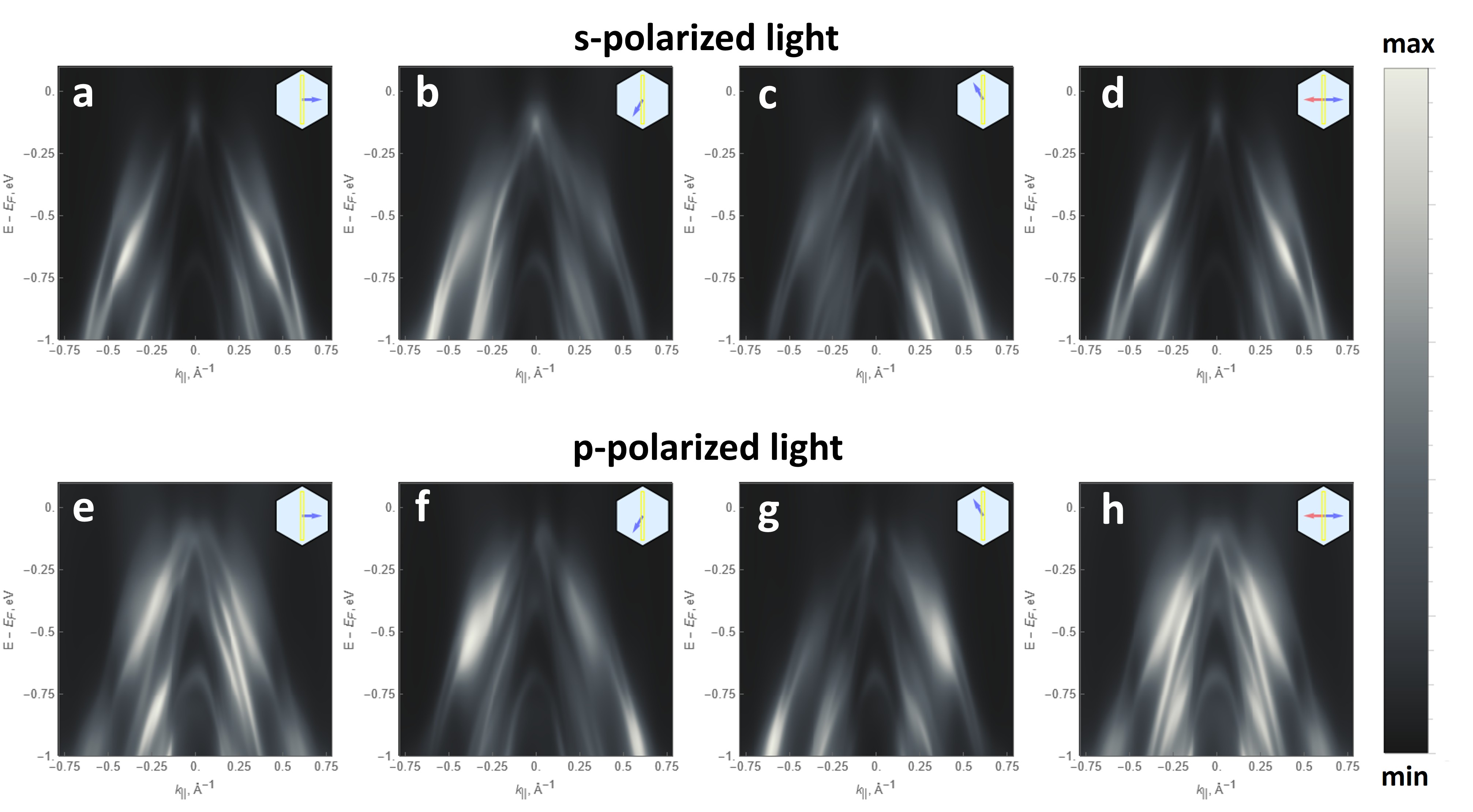}
    \caption{\label{figure: 1-step SI}Same as fig. \ref{figure: 1-step}, but displaying spin-integrated photoelectron current: Calculated ARPES bandmaps at the $k_z=0$ nodal plane at $h\nu=82$~eV along $\overline{\text{K}}-\overline{\Gamma}-\overline{\text{K}}$ direction. The domain orientations and experimental geometry in each plot are sketched in the inset. (a)-(d) correspond to the $s$-polarized light, (e)-(h) correspond to $p$-polarized light.}
\end{figure}

\begin{figure}[h]
\centering
    \includegraphics[width=0.95\linewidth]{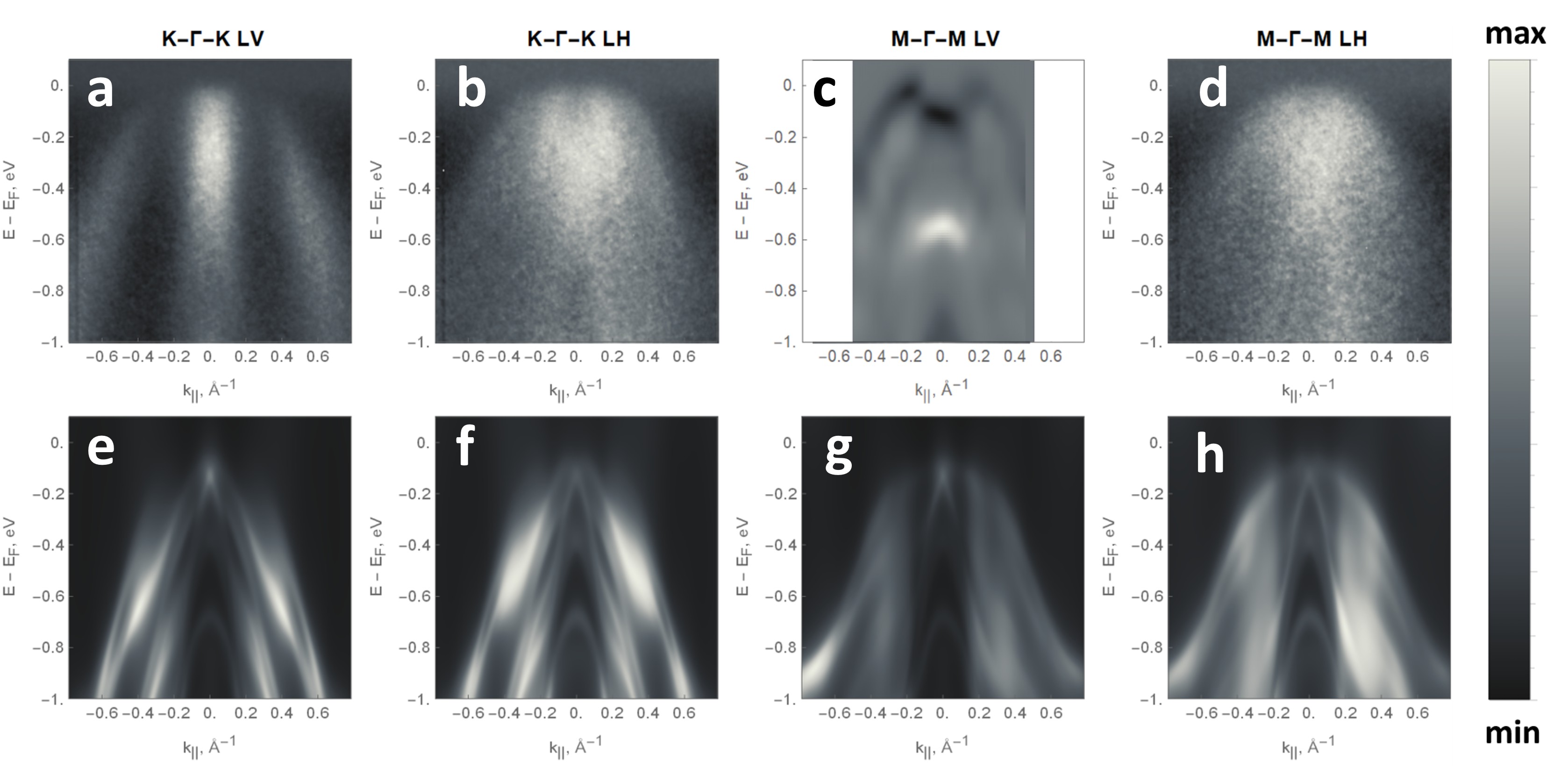}
\caption{\label{figure: bandmaps int}Measured (a-d) and 1-step calculated (e-h) spin-integrated band maps for the fully-compensated domain configuration. The light polarization and reciprocal space directions are indicated. The experimental data in (a), (b) and (d) is shown with angle-integrated background subtracted, and the data in (c) is displayed as a curvature plot for better visualization.}
\end{figure}

Figures \ref{figure: 1-step SI} and \ref{figure: bandmaps int} focus on spin-integrated 1-step photoemission calculations and experimental ARPES data. The calculations predict a strong dependence of photoelectron distribution on the N\'eel vector direction and light polarization. However, due to the finite energy and angular resolution in the experiment, we were not able to employ this effect to access the domain configuration under the photon beam. Figure \ref{figure: bandmaps int} displays a comparison between spin-integrated ARPES bandmaps and the ones calculated for a domain combination, where all opposite N\'eel vectors are equally present and thus spin-compensated. Due to the p-doping of the sample, the calculated bands are shifted by 100 meV compared to those measured by ARPES. 

In addition to the energy shift, a slight mismatch in $k_{||}$ scale of around 15\% is also present. We hypothesize that this discrepancy may come either from slight detuning of the $U$ parameter in the DFT+U calculations (PLEASE COMMENT ON THIS MORE EXPLICITLY), or from a small misalignment of $k_z$. Apart from these differences, a good qualitative agreement is evident.

\section{\label{sec:Impact of matrix element effects on spin polarization of photoelectrons}Impact of matrix element effects on spin polarization of photoelectrons}

\begin{figure}[H]
\centering
    \includegraphics[width=0.95\linewidth]{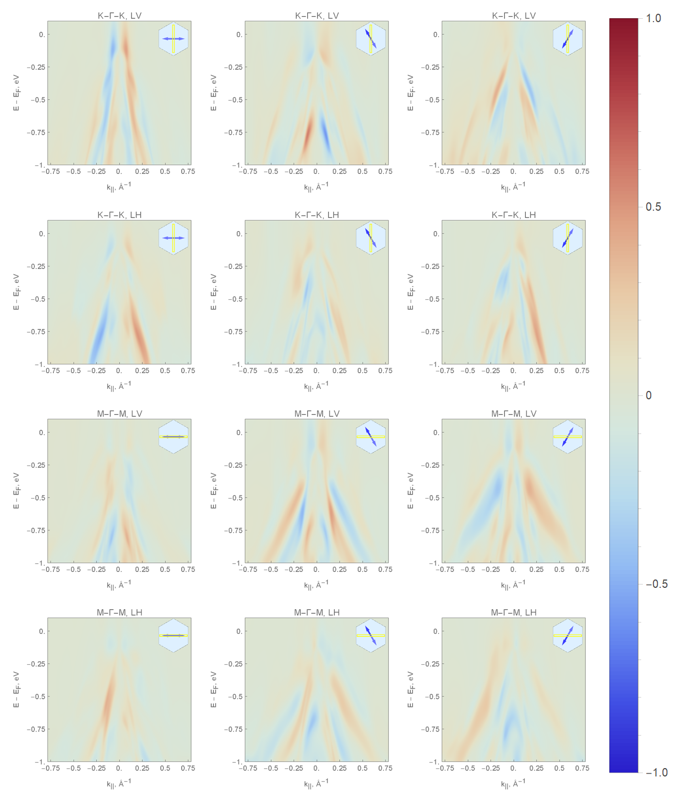}
    \caption{\label{figure: PZ MEE grid}Bandmaps of spin polarization arising from the photoemission process for differently oriented domains and $s$- and $p$- polarized light (LV and LH, respectively).}
\end{figure}

In order to deduct the contribution of the matrix element effects on the observed spin polarization, we performed the following analysis of the present 1-step photoemission calculations. By using the same ground state spectral function, we computed difference in SARPES signal between two opposite incidence angles of the photon beam. This is equivalent to a scenario, when the light incidence is kept constant, but the sample as a whole undergoes azimuthal rotation by 180 degrees. Thus, the difference between these two orientations highlights spectral features, which change sign upon the sample rotation and hence are attributed only to the effect of geometry in matrix elements of photoemission. The results for $\overline{\text{K}}-\overline{\Gamma}-\overline{\text{K}}$ and $\overline{\text{M}}-\overline{\Gamma}-\overline{\text{M}}$ with $s$- and $p$-polarized light are displayed in Fig. \ref{figure: PZ MEE grid}. We emphasize that 180$^o$ azimuthal rotation of the sample is not the same as flipping the N\'eel vector, since the first one does not change the ground-state spin polarization of the band structure, but the second one causes a sign flip of $S_z$.

The SARPES signal presented below originates from the geometrical contribution of matrix element effects and will be always present on top of features related to the ground-state polarization. Noteworthy, this contribution is purely anti-symmetric with respect to the $\Gamma$ point for the $s$-polarized light (LV), when the electric field vector $\vv{E}$ always lies in the sample's surface plane. On the contrary, when $\vv{E}$ has a finite angle with respect to the surface (in our case - 45 degrees), some symmetric part of the photoemission-induced spin signal is also present. In addition, we indicate a minimal impact in the region-of-interest - $k_{||} > 0.25$ and $E_B < 0.5 eV$. This region is hence a good choice to access spin textures related to the ground state. 

\section{\label{sec:Circular dichroism in ARPES data}Circular dichroism in ARPES data}

\begin{figure}[H]
\centering
    \includegraphics[width=0.9\linewidth]{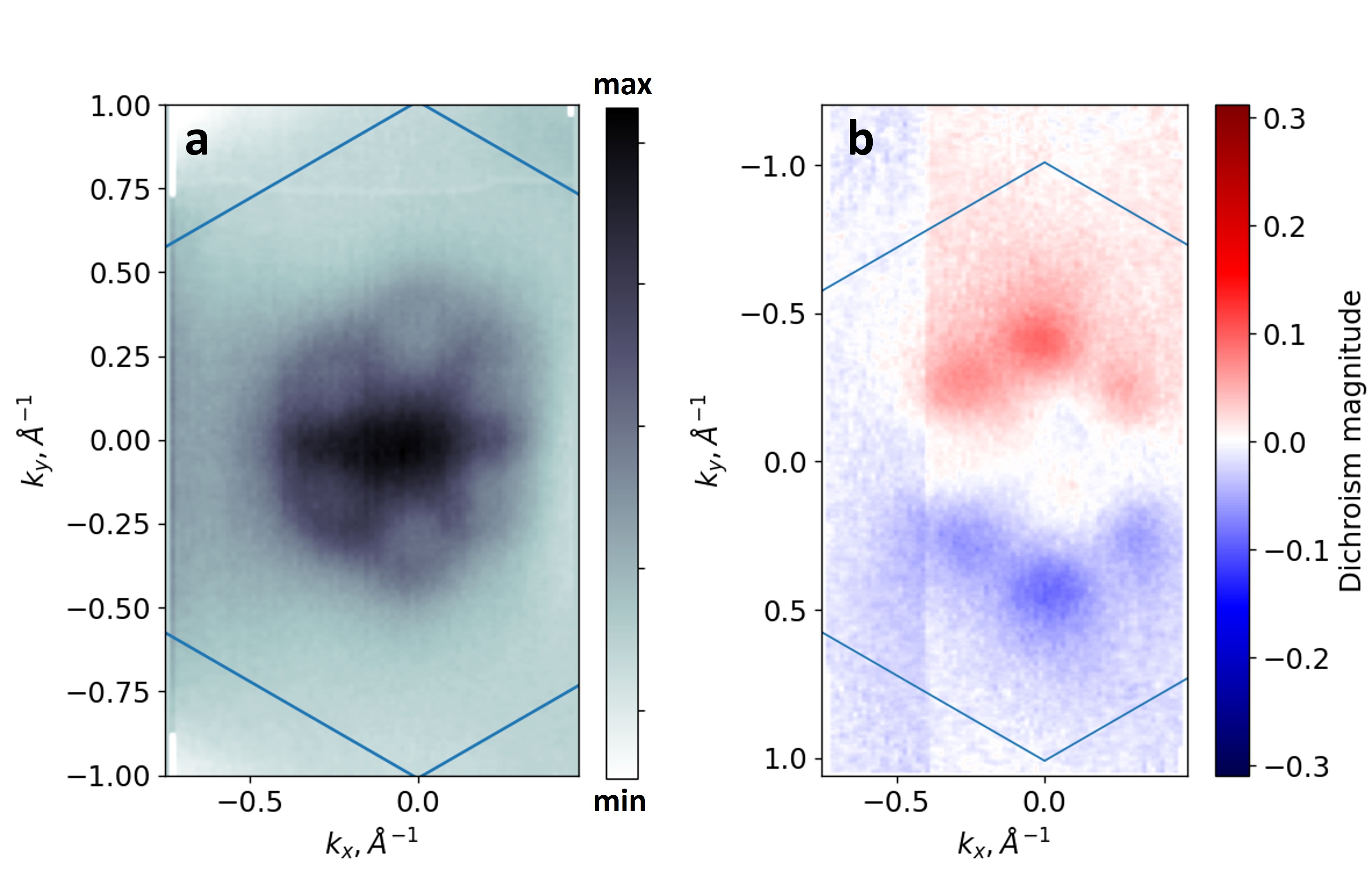}
    \caption{\label{figure: CDAD}Constant energy cuts of circular integrated (a) and circular dichroic (b) signal from a non-field-cooled sample.}
\end{figure}

In Fig. \ref{figure: CDAD}(a) we show constant energy cuts obtained by summing signal from both circular light polarizations from a non-field-cooled sample. So-called circular integrated signal effectively suppress the contribution of matrix element effects in angular distribution of photoelectrons. The angular distribution of the spectral weight shows predominant 6-fold symmetry. Because we do not observe 2-fold signatures of a single domain, it does not seem possible to deduct the exact domain configuration, taking into account finite energy and angular resolution of the ARPES experiment with respect to the spin splitting difference between nonequivalent high-symmetry directions. 

The observed pattern of circular dichroism in angular distribution of photoelectrons (CDAD) in Fig. \ref{figure: CDAD}(b) is antisymmetric with respect to the scattering plane of photoemission (the plane formed by the direction of the incident light and the surface normal). Since it coincides with the mirror plane of the non-magnetic crystal, we fully attribute the observed CDAD to the geometry of the photoemission experiment \cite{schonhense1990circular, vidal2007circular}. Therefore, it suggests that spin-integrated ARPES is much less sensitive to the domain orientation than spin-resolved ARPES. 

\section{\label{sec:Supplementary SARPES data}Supplementary SARPES data}

\begin{figure}[H]
\centering
    \includegraphics[width=0.95\linewidth]{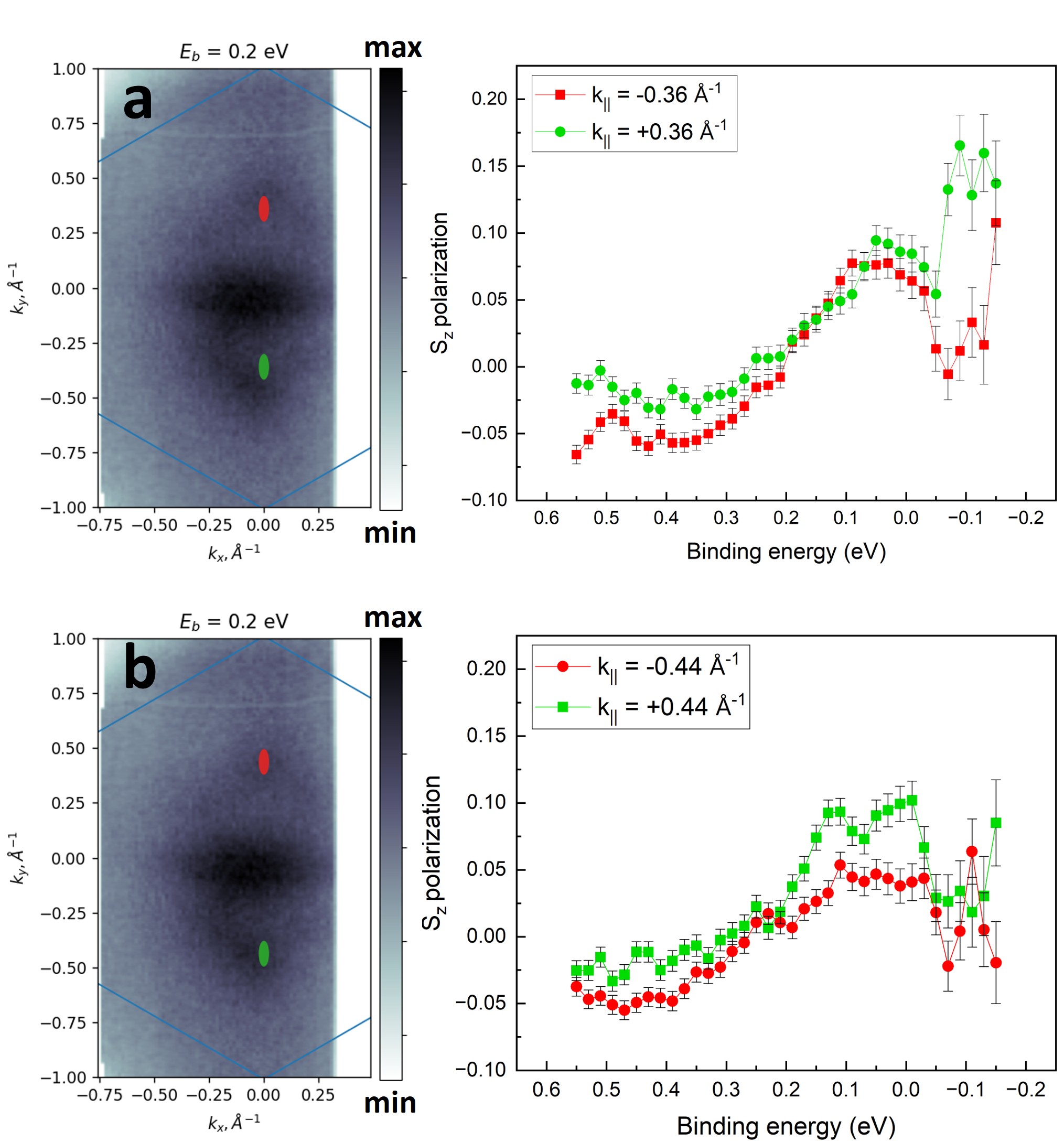}
    \caption{\label{figure: SARPES SI}SARPES data in more detail: spin-resolved EDCs and their angular position illustrated on iso-energy cuts at $E_B = 0.2~eV$. Patches (red and green) illustrate angular integration windows parallel and perpendicular to the HSA slits.}
\end{figure}

\pagebreak

\bibliographystyle{unsrt}
\bibliography{references}

@article{schoenhense2022,
    author = {Schönhense, Gerd and Elmers, Hans-Joachim},
    title = {Spin- and time-resolved photoelectron spectroscopy and diffraction studies using time-of-flight momentum microscopes},
    journal = {Journal of Vacuum Science and Technology A},
    volume = {40},
    number = {2},
    pages = {020802},
    year = {2022},
    month = {01},
    issn = {0734-2101},
    doi = {10.1116/6.0001500},
    url = {https://doi.org/10.1116/6.0001500},
    eprint = {https://pubs.aip.org/avs/jva/article-pdf/doi/10.1116/6.0001500/19811654/020802_1_online.pdf},
}

@article{vsmejkal2022emerging,
  title={Emerging research landscape of altermagnetism},
  author={{\v{S}}mejkal, Libor and Sinova, Jairo and Jungwirth, Tomas},
  journal={Physical Review X},
  volume={12},
  number={4},
  pages={040501},
  year={2022},
  publisher={APS}
}

@article{jungwirth2026symmetry,
  title={Symmetry, microscopy and spectroscopy signatures of altermagnetism},
  author={Jungwirth, Tomas and Sinova, Jairo and Fernandes, Rafael M and Liu, Qihang and Watanabe, Hikaru and Murakami, Shuichi and Nakatsuji, Satoru and {\v{S}}mejkal, Libor},
  journal={Nature},
  volume={649},
  number={8098},
  pages={837--847},
  year={2026},
  publisher={Nature Publishing Group UK London}
}

@article{campos2026persistent,
  title={Persistent altermagnetism},
  author={Campos, Warlley H and Mbognou, FC and Hellenes, Anna Birk and Poata, Joseph and Chen, Taikang and Priessnitz, Jan and {\v{S}}mejkal, Libor},
  journal={arXiv preprint arXiv:2603.12223},
  year={2026}
}

@article{vsmejkal2022beyond,
  title={Beyond conventional ferromagnetism and antiferromagnetism: A phase with nonrelativistic spin and crystal rotation symmetry},
  author={{\v{S}}mejkal, Libor and Sinova, Jairo and Jungwirth, Tomas},
  journal={Physical Review X},
  volume={12},
  number={3},
  pages={031042},
  year={2022},
  publisher={APS}
}

@article{song2025altermagnets,
  title={Altermagnets as a new class of functional materials},
  author={Song, Cheng and Bai, Hua and Zhou, Zhiyuan and Han, Lei and Reichlova, Helena and Dil, J Hugo and Liu, Junwei and Chen, Xianzhe and Pan, Feng},
  journal={Nature Reviews Materials},
  pages={1--13},
  year={2025},
  publisher={Nature Publishing Group UK London}
}

@article{krempasky2024altermagnetic,
  title={Altermagnetic lifting of Kramers spin degeneracy},
  author={Krempask{\`y}, Juraj and {\v{S}}mejkal, L and D’souza, SW and Hajlaoui, M and Springholz, G and Uhl{\'\i}{\v{r}}ov{\'a}, K and Alarab, F and Constantinou, PC and Strocov, V and Usanov, D and others},
  journal={Nature},
  volume={626},
  number={7999},
  pages={517--522},
  year={2024},
  publisher={Nature Publishing Group UK London}
}

@article{lee2024broken,
  title={Broken Kramers degeneracy in altermagnetic MnTe},
  author={Lee, Suyoung and Lee, Sangjae and Jung, Saegyeol and Jung, Jiwon and Kim, Donghan and Lee, Yeonjae and Seok, Byeongjun and Kim, Jaeyoung and Park, Byeong Gyu and {\v{S}}mejkal, Libor and others},
  journal={Physical review letters},
  volume={132},
  number={3},
  pages={036702},
  year={2024},
  publisher={APS}
}

@article{osumi2024observation,
  title={Observation of a giant band splitting in altermagnetic MnTe},
  author={Osumi, T and Souma, S and Aoyama, T and Yamauchi, K and Honma, A and Nakayama, K and Takahashi, T and Ohgushi, Kenya and Sato, Takafumi},
  journal={Physical Review B},
  volume={109},
  number={11},
  pages={115102},
  year={2024},
  publisher={APS}
}

@phdthesis{lee2024investigations,
  title={Investigations on the electronic structures of magnetic transition metal chalcogenides},
  author={Suyoung Lee},
  year={2024},
  school={Department of Physics and Astronomy, Graduate School, Seoul National University}
}

@article{martuza2025itinerant,
  title={Itinerant and Correlated Nature of Altermagnetic MnTe Single Crystal Studied by Photoemission and Inverse-Photoemission Spectroscopies},
  author={Martuza, Kazi Golam and Kumar, Yogendra and Yamaguchi, Hiroshi and Kumar, Shiv and Arita, Masashi and Sato, Hitoshi and Ideta, Shin-ichiro and Shimada, Kenya},
  journal={Materials},
  volume={18},
  number={13},
  pages={3103},
  year={2025},
  publisher={MDPI}
}

@article{hajlaoui2024temperature,
    author = {Hajlaoui, Mahdi and Wilfred D'Souza, Sunil and Šmejkal, Libor and Kriegner, Dominik and Krizman, Gauthier and Zakusylo, Tetiana and Olszowska, Natalia and Caha, Ondřej and Michalička, Jan and Sánchez-Barriga, Jaime and Marmodoro, Alberto and Výborný, Karel and Ernst, Arthur and Cinchetti, Mirko and Minar, Jan and Jungwirth, Tomas and Springholz, Gunther},
    title = {Temperature Dependence of Relativistic Valence Band Splitting Induced by an Altermagnetic Phase Transition},
    journal = {Advanced Materials},
    volume = {36},
    number = {31},
    pages = {2314076},
    doi = {https://doi.org/10.1002/adma.202314076},
    url = {https://onlinelibrary.wiley.com/doi/abs/10.1002/adma.202314076},
    eprint = {https://onlinelibrary.wiley.com/doi/pdf/10.1002/adma.202314076},
    year = {2024}
}

@article{lee2025dichotomous,
  title={Dichotomous Temperature Response in the Electronic Structure of Epitaxially Grown Altermagnet MnTe},
  author={Lee, Ji-Eun and Zhong, Yong and Li, Qile and Edmonds, Mark T and Shen, Zhi-Xun and Hwang, Choongyu and Mo, Sung-Kwan},
  journal={Nano Letters},
  year={2025},
  publisher={ACS Publications}
}

@article{amin2024nanoscale,
  title={Nanoscale imaging and control of altermagnetism in MnTe},
  author={Amin, OJ and Dal Din, A and Golias, E and Niu, Y and Zakharov, A and Fromage, SC and Fields, CJB and Heywood, SL and Cousins, RB and Maccherozzi, F and others},
  journal={Nature},
  volume={636},
  number={8042},
  pages={348--353},
  year={2024},
  publisher={Nature Publishing Group}
}

@article{orlova2025magnetization,
  title={Magnetization symmetry for the MnTe altermagnetic candidate},
  author={Orlova, NN and Esin, VD and Timonina, AV and Kolesnikov, NN and Deviatov, EV},
  journal={arXiv preprint arXiv:2502.11876},
  year={2025}
}

@article{orlova2024crossover,
  title={Crossover from relativistic to non-relativistic net magnetization for MnTe altermagnet candidate},
  author={Orlova, Nadezhda Nikolaevna and Avakyants, Artem Aleksandrovich and Timonina, Anna Vladimirovna and Kolesnikov, Nikolay Nikolaevich and Deviatov, EV},
  journal={JETP Letters},
  volume={120},
  number={5},
  pages={360--366},
  year={2024},
  publisher={Springer}
}

@article{belashchenko2025giant,
  title={Giant strain-induced spin splitting effect in MnTe, ag-wave altermagnetic semiconductor},
  author={Belashchenko, KD},
  journal={Physical Review Letters},
  volume={134},
  number={8},
  pages={086701},
  year={2025},
  publisher={APS}
}

@article{schlesinger1998mn,
  title={The Mn-Te (manganese-tellurium) system},
  author={Schlesinger, Mark E},
  journal={Journal of phase equilibria},
  volume={19},
  number={6},
  pages={591--596},
  year={1998},
  publisher={Springer}
}

@article{jain2024buffer,
  title={Buffer-layer-controlled nickeline vs zinc-blende/wurtzite-type MnTe growths on c-plane Al 2 O 3 substrates},
  author={Jain, Deepti and Yi, Hee Taek and Mazza, Alessandro R and Kisslinger, Kim and Han, Myung-Geun and Brahlek, Matthew and Oh, Seongshik},
  journal={Physical Review Materials},
  volume={8},
  number={1},
  pages={014203},
  year={2024},
  publisher={APS}
}

@article{bossini2020exchange,
  title={Exchange-mediated magnetic blue-shift of the band-gap energy in the antiferromagnetic semiconductor MnTe},
  author={Bossini, Davide and Terschanski, Marc and Mertens, Fabian and Springholz, Gunther and Bonanni, Alberta and Uhrig, G{\"o}tz S and Cinchetti, Mirko},
  journal={New Journal of Physics},
  volume={22},
  number={8},
  pages={083029},
  year={2020},
  publisher={IOP Publishing}
}

@article{kriegner2016multiple,
  title={Multiple-stable anisotropic magnetoresistance memory in antiferromagnetic MnTe},
  author={Kriegner, Dominik and V{\`y}born{\`y}, K and Olejn{\'\i}k, K and Reichlov{\'a}, H and Nov{\'a}k, V and Marti, X and Gazquez, J and Saidl, V and N{\v{e}}mec, P and Volobuev, VV and others},
  journal={Nature communications},
  volume={7},
  number={1},
  pages={11623},
  year={2016},
  publisher={Nature Publishing Group UK London}
}

@article{gonzalez2023spontaneous,
  title={Spontaneous anomalous Hall effect arising from an unconventional compensated magnetic phase in a semiconductor},
  author={Gonzalez Betancourt, RD and Zub{\'a}{\v{c}}, Jan and Gonzalez-Hernandez, R and Geishendorf, Kevin and {\v{S}}ob{\'a}{\v{n}}, Zbynek and Springholz, Gunther and Olejn{\'\i}k, Kamil and {\v{S}}mejkal, Libor and Sinova, Jairo and Jungwirth, Tomas and others},
  journal={Physical Review Letters},
  volume={130},
  number={3},
  pages={036702},
  year={2023},
  publisher={APS}
}

@article{hariki2024x,
  title={X-ray magnetic circular dichroism in altermagnetic $\alpha$-MnTe},
  author={Hariki, A and Dal Din, A and Amin, OJ and Yamaguchi, T and Badura, A and Kriegner, D and Edmonds, KW and Campion, RP and Wadley, P and Backes, D and others},
  journal={Physical Review Letters},
  volume={132},
  number={17},
  pages={176701},
  year={2024},
  publisher={APS}
}

@article{takegami2025circular,
  title={Circular Dichroism in Resonant Inelastic X-ray Scattering: Probing Altermagnetic Domains in MnTe},
  author={Takegami, D and Aoyama, T and Okauchi, T and Yamaguchi, T and Tippireddy, S and Agrestini, S and Garc{\'\i}a-Fern{\'a}ndez, M and Mizokawa, T and Ohgushi, K and Zhou, Ke-Jin and others},
  journal={arXiv preprint arXiv:2502.10809},
  year={2025}
}

@article{kruger2025circular,
  title={Circular dichroism in resonant photoelectron diffraction as a direct probe of sublattice magnetization in altermagnets},
  author={Kr{\"u}ger, Peter},
  journal={arXiv preprint arXiv:2504.08380},
  year={2025}
}

@article{lovesey2023templates,
  title={Templates for magnetic symmetry and altermagnetism in hexagonal MnTe},
  author={Lovesey, SW and Khalyavin, DD and Van Der Laan, G},
  journal={Physical Review B},
  volume={108},
  number={17},
  pages={174437},
  year={2023},
  publisher={APS}
}

@article{hariki2024determination,
  title={Determination of the N{\'e}el vector in rutile altermagnets through x-ray magnetic circular dichroism: The case of MnF 2},
  author={Hariki, A and Okauchi, T and Takahashi, Y and Kune{\v{s}}, J},
  journal={Physical Review B},
  volume={110},
  number={10},
  pages={L100402},
  year={2024},
  publisher={APS}
}

@article{kresse1996efficient,
  title={Efficient iterative schemes for ab initio total-energy calculations using a plane-wave basis set},
  author={Kresse, Georg and Furthm{\"u}ller, J{\"u}rgen},
  journal={Physical review B},
  volume={54},
  number={16},
  pages={11169},
  year={1996},
  publisher={APS}
}

@article{moser2017experimentalist,
  title={An experimentalist's guide to the matrix element in angle resolved photoemission},
  author={Moser, Simon},
  journal={Journal of Electron Spectroscopy and Related Phenomena},
  volume={214},
  pages={29--52},
  year={2017},
  publisher={Elsevier}
}

@article{riley2014direct,
  title={Direct observation of spin-polarized bulk bands in an inversion-symmetric semiconductor},
  author={Riley, Jonathon Mark and Mazzola, F and Dendzik, M and Michiardi, M and Takayama, T and Bawden, Lewis and Graner{\o}d, C and Leandersson, Mats and Balasubramanian, T and Hoesch, M and others},
  journal={Nature Physics},
  volume={10},
  number={11},
  pages={835--839},
  year={2014},
  publisher={Nature Publishing Group UK London}
}

@article{polley2024bloch,
  title={The Bloch Beamline at MAX IV: Micro-Spot ARPES from a Conventional, Full-Featured Beamline},
  author={Polley, CM and Leandersson, M and Adell, J and Osiecki, J and Carbone, D and Ali, K and Fedderwitz, H and Balasubramanian, T},
  journal={Synchrotron Radiation News},
  volume={37},
  number={4},
  pages={18--23},
  year={2024},
  publisher={Taylor \& Francis}
}

@article{okuda2017recent,
  title={Recent trends in spin-resolved photoelectron spectroscopy},
  author={Okuda, Taichi},
  journal={Journal of Physics: Condensed Matter},
  volume={29},
  number={48},
  pages={483001},
  year={2017},
  publisher={IOP Publishing}
}

@article{dil2019spin,
  title={Spin-and angle-resolved photoemission on topological materials},
  author={Dil, J Hugo},
  journal={Electronic Structure},
  volume={1},
  number={2},
  pages={023001},
  year={2019},
  publisher={IOP Publishing}
}

@article{heinzmann2012spin,
  title={Spin--orbit-induced photoelectron spin polarization in angle-resolved photoemission from both atomic and condensed matter targets},
  author={Heinzmann, Ulrich and Dil, J Hugo},
  journal={Journal of Physics: Condensed Matter},
  volume={24},
  number={17},
  pages={173001},
  year={2012},
  publisher={IOP Publishing}
}

@article{meier2011interference,
  title={Interference of spin states in photoemission from Sb/Ag (111) surface alloys},
  author={Meier, Fabian and Petrov, Vladimir and Mirhosseini, Hossein and Patthey, Luc and Henk, J{\"u}rgen and Osterwalder, J{\"u}rg and Dil, J Hugo},
  journal={Journal of Physics: Condensed Matter},
  volume={23},
  number={7},
  pages={072207},
  year={2011},
  publisher={IOP Publishing}
}

@article{fanciulli2017spin,
  title={Spin polarization and attosecond time delay in photoemission from spin degenerate states of solids},
  author={Fanciulli, Mauro and Volfov{\'a}, Henrieta and Muff, Stefan and Braun, J{\"u}rgen and Ebert, Hubert and Min{\'a}r, Jan and Heinzmann, Ulrich and Dil, J Hugo},
  journal={Physical review letters},
  volume={118},
  number={6},
  pages={067402},
  year={2017},
  publisher={APS}
}

@article{vsmejkal2022anomalous,
  title={Anomalous hall antiferromagnets},
  author={{\v{S}}mejkal, Libor and MacDonald, Allan H and Sinova, Jairo and Nakatsuji, Satoru and Jungwirth, Tomas},
  journal={Nature Reviews Materials},
  volume={7},
  number={6},
  pages={482--496},
  year={2022},
  publisher={Nature Publishing Group UK London}
}

@phdthesis{usanov2025, 
    address={Lausanne}, 
    title={Spin-resolved electron spectroscopy: towards new instrumentation and unconventional magnetism}, 
    url={https://infoscience.epfl.ch/handle/20.500.14299/247563}, 
    DOI={10.5075/epfl-thesis-11240}, 
    school={EPFL}, 
    author={Usanov Dmitrii}, 
    year={2025}
}

@article{pescia1985determination,
  title={Determination of observable conduction band symmetry in angle-resolved electron spectroscopies: Non-symmorphic space groups},
  author={Pescia, D and Law, AR and Johnson, MT and Hughes, HP},
  journal={Solid state communications},
  volume={56},
  number={9},
  pages={809--812},
  year={1985},
  publisher={Elsevier}
}

@article{prince1986symmetry,
  title={The symmetry-based constraints in angle-resolved photoemission from structures belonging to non-symmorphic space groups: p (2$\times$ 2)- CNi (100)},
  author={Prince, KC and Surman, M and Lindner, Th and Bradshaw, AM},
  journal={Solid state communications},
  volume={59},
  number={2},
  pages={71--75},
  year={1986},
  publisher={Elsevier}
}

@article{ryoo2018momentum,
  title={Momentum-dependent spin selection rule in photoemission with glide symmetry},
  author={Ryoo, Ji Hoon and Park, Cheol-Hwan},
  journal={Physical Review B},
  volume={98},
  number={23},
  pages={235403},
  year={2018},
  publisher={APS}
}

@article{alexandradinata2020glide,
  title={Glide-resolved photoemission spectroscopy: Measuring topological invariants in nonsymmorphic space groups},
  author={Alexandradinata, Aris and Wang, Zhijun and Bernevig, B Andrei and Zaletel, Michael},
  journal={Physical Review B},
  volume={101},
  number={23},
  pages={235166},
  year={2020},
  publisher={APS}
}

@article{arpiainen2006effect,
  title={Effect of Symmetry Distortions on Photoelectron Selection Rules<? format?> and Spectra of Bi 2 Sr 2 CaCu 2 O 8+ $\delta$},
  author={Arpiainen, V and Lindroos, M},
  journal={Physical review letters},
  volume={97},
  number={3},
  pages={037601},
  year={2006},
  publisher={APS}
}

@article{matsui20184,
  title={The 4 $\pi$ kz periodicity in photoemission from graphite},
  author={Matsui, Fumihiko and Nishikawa, Hiroaki and Daimon, Hiroshi and Muntwiler, Matthias and Takizawa, Masaru and Namba, Hidetoshi and Greber, Thomas},
  journal={Physical Review B},
  volume={97},
  number={4},
  pages={045430},
  year={2018},
  publisher={APS}
}

@article{din2025unconventional,
  title={Unconventional relativistic spin polarization of electronic bands in an altermagnet},
  author={Din, A Dal and Usanov, DA and {\v{S}}mejkal, L and D'Souza, SW and Guo, F and Amin, OJ and Dawa, EM and Campion, RP and Edmonds, KW and Kiraly, B and others},
  journal={arXiv preprint arXiv:2511.01690},
  year={2025}
}

@article{zeng2025non,
  title={Non-altermagnetic spin texture in MnTe},
  author={Zeng, Meng and Liu, Pengfei and Zhu, Ming-Yuan and Zheng, Naifu and Liu, Xiang-Rui and Zhu, Yu-Peng and Shao, Tian-Hao and Hao, Yu-Jie and Ma, Xiao-Ming and Qu, Gexing and others},
  journal={arXiv preprint arXiv:2511.02447},
  year={2025}
}

@article{kriegner2017magnetic,
  title={Magnetic anisotropy in antiferromagnetic hexagonal MnTe},
  author={Kriegner, D and Reichlova, H and Grenzer, J and Schmidt, W and Ressouche, E and Godinho, J and Wagner, T and Martin, SY and Shick, AB and Volobuev, VV and others},
  journal={Physical Review B},
  volume={96},
  number={21},
  pages={214418},
  year={2017},
  publisher={APS}
}

@article{perdew1996generalized,
  title={Generalized gradient approximation made simple},
  author={Perdew, John P and Burke, Kieron and Ernzerhof, Matthias},
  journal={Physical review letters},
  volume={77},
  number={18},
  pages={3865},
  year={1996},
  publisher={APS}
}

@article{pendry1976theory,
  title={Theory of photoemission},
  author={Pendry, JB},
  journal={Surface Science},
  volume={57},
  number={2},
  pages={679--705},
  year={1976},
  publisher={Elsevier}
}

@article{hopkinson1980calculation,
  title={Calculation of photoemission spectra for surfaces of solids},
  author={Hopkinson, JFL and Pendry, JB and Titterington, DJ},
  journal={Computer Physics Communications},
  volume={19},
  number={1},
  pages={69--92},
  year={1980},
  publisher={Elsevier}
}

@incollection{pendry1974low,
  title={Low-energy electron diffraction},
  author={Pendry, John Brian},
  booktitle={Interaction of Atoms and Molecules with Solid Surfaces},
  pages={201--211},
  year={1974},
  publisher={Springer}
}

@misc{ebert2012munich,
  title={The Munich SPR-KKR Package},
  author={Ebert, H and others},
  year={2012},
  publisher={version 9.4},
  url={http://olymp.cup.uni-muenchen.de/ak/ebert/SPRKKR}, 
}

@article{Ebert_2011,
    doi = {10.1088/0034-4885/74/9/096501},
    url = {https://doi.org/10.1088/0034-4885/74/9/096501},
    year = {2011},
    month = {aug},
    publisher = {},
    volume = {74},
    number = {9},
    pages = {096501},
    author = {Ebert, H and Ködderitzsch, D and Minár, J},
    title = {Calculating condensed matter properties using the KKR-Green's function method—recent developments and applications},
    journal = {Reports on Progress in Physics}
}

@article{gray2011probing,
  title={Probing bulk electronic structure with hard X-ray angle-resolved photoemission},
  author={Gray, AX and Papp, C and Ueda, Shigenori and Balke, B and Yamashita, Y and Plucinski, L and Min{\'a}r, J and Braun, J and Ylvisaker, ER and Schneider, CM and others},
  journal={Nature materials},
  volume={10},
  number={10},
  pages={759--764},
  year={2011},
  publisher={Nature Publishing Group UK London}
}

@article{malmstrom1980program,
  title={A program for calculation of the reflection and transmission of electrons through a surface potential barrier},
  author={Malmstr{\"o}m, G and Rundgren, J},
  journal={Computer Physics Communications},
  volume={19},
  number={2},
  pages={263--270},
  year={1980},
  publisher={Elsevier}
}

@article{braun1996theory,
  title={The theory of angle-resolved ultraviolet photoemission and its applications to ordered materials},
  author={Braun, J},
  journal={Reports on Progress in Physics},
  volume={59},
  number={10},
  pages={1267--1338},
  year={1996}
}

@article{braun2018correlation,
  title={Correlation, temperature and disorder: Recent developments in the one-step description of angle-resolved photoemission},
  author={Braun, J{\"u}rgen and Min{\'a}r, J{\'a}n and Ebert, Hubert},
  journal={Physics Reports},
  volume={740},
  pages={1--34},
  year={2018},
  publisher={Elsevier}
}

@article{schonhense1990circular,
  title={Circular dichroism and spin polarization in photoemission from adsorbates and non-magnetic solids},
  author={Sch{\"o}nhense, Gerd},
  journal={Physica Scripta},
  volume={1990},
  number={T31},
  pages={255--275},
  year={1990}
}

@article{vidal2007circular,
  title={Circular dichroism in photoemission as a fingerprint of surface band structure: The case of ZnSe (001)-c (2$\times$ 2)},
  author={Vidal, Franck and Marangolo, M and Torelli, P and Eddrief, M and Mulazzi, M and Panaccione, G},
  journal={Physical Review B—Condensed Matter and Materials Physics},
  volume={76},
  number={8},
  pages={081302},
  year={2007},
  publisher={APS}
}

@article{hicken2025anomalous,
  title={Anomalous temperature dependence of local magnetic fields in altermagnetic MnTe},
  author={Hicken, Thomas J and Amin, Oliver and Din, Alfred Dal and Kriegner, Dominik and Luetkens, Hubertus and Reichlov{\'a}, Helena and Salman, Zaher and Uhl{\'\i}{\v{r}}ov{\'a}, Kl{\'a}ra and Wadley, Peter and Krempask{\`y}, Juraj and others},
  journal={arXiv preprint arXiv:2507.14710},
  year={2025}
}

\end{document}